\documentclass[superscriptaddress]{emulateapj}
\usepackage{amsmath}

\shorttitle{X-Ray Feedback from AGN}

\begin{document}
\title{The Effects of X-Ray Feedback from AGN on Host Galaxy Evolution}
\author{D. Clay Hambrick}
\author{Jeremiah P. Ostriker}
\affil{Princeton University Observatory, Princeton, NJ 08544}
\author{Thorsten Naab}
\affil{Max-Planck-Insitut f\"ur Astrophysik, Karl-Schwarzschild-Str.\ 1,
D-85741, Garching bei M\"unchen, Germany \\}
\author{Peter H. Johansson}
\affil{Finnish Centre for  Astronomy with ESO, University of Turku, 
V\"ais\"al\"antie 20, FI-21500 Piikki\"o, Finland\\
Department of Physics, University of Helsinki, Gustaf H\"allstr\"omin 
katu 2a, FI-00014 Helsinki, Finland \\
 Universit\"ats-Sternwarte M\"unchen, Scheinerstr.\ 1, D-81679 M\"unchen, 
Germany \\}

\begin{abstract}
Hydrodynamic simulations of galaxies with active galactic nuclei (AGN) have
typically employed feedback that 
is purely local: i.e., an injection of energy to the
immediate neighborhood of the black hole.  We perform  GADGET-2 simulations of
massive elliptical galaxies with an
additional 
feedback component: an observationally calibrated X-ray radiation
field which emanates from 
the black hole and heats gas out to large radii from the galaxy
center.  We find that including the heating and radiation
pressure associated with this X-ray flux in our simulations enhances
the effects which are 
commonly reported from AGN feedback. This new
feedback model is twice as effective as traditional feedback at
suppressing star formation, produces 3 times less
star formation in the last 6 Gyr, and modestly lowers the final BH
mass (30\%). It is also significantly more effective than an X-ray
background in 
reducing the number of satellite galaxies.  
\end{abstract}
\maketitle

\section{Introduction}
It has been well established that massive galaxies typically
contain massive black holes (BHs) at their centers, and it is widely
believed that these black holes, while in their
so-called ``active galactic nuclei'' (AGN) phases, can have profound
influence on 
the galaxies which host them (see \citealt{ff05} for
a comprehensive observational review).  This influence is inferred from,
among other observations, the correlation between the BH mass and
the bulge mass of the host galaxy, which was first noticed by \citet{kr95} and
studied by \citet{mag98}, and has
since been repeatedly confirmed \citep{hr04}, as has the somewhat tighter
relation between the BH mass and the host bulge's velocity dispersion,
discovered by \citet{fm00} and \citet{geb00}. The inferred BH mass
distribution has also been convincingly linked to the quasar
luminosity function 
\citep{yt02}.

Since the work of \citet{sr98}, the
relationship between the AGN and host galaxy has generally been
characterized as a process of feedback: the galaxy supplies gas
to be accreted by the BH, which emits some fraction of the accreted
mass as mechanical energy to the surroundings, some fraction via the
broad-line winds, and some fraction as the radio jets.  All these
processes can heat the surrounding gas, and since gas
which has been heated is less dense and thus accretes more slowly, the
feedback process is 
described as ``self-regulating'' (i.e. a negative feedback loop).

Thus there are two modes by which AGN may give energy
to their surroundings: the ``mechanical'' mode, where the AGN
inflates local gas bubbles which expand through the inter-stellar and
inter-galactic media (ISM/IGM) \citep{fab00,
  chur02, self05}; and the ``electromagnetic'' mode, where photons
from the AGN accretion region for which the neighboring gas has a low
optical depth escape and directly interact with the rest of the ISM/IGM
\citep{socs05}.  Both of these modes may then interact with the
surrounding galactic gas by either (or rather, both) of two
mechanisms:  energy-based (i.e.\ heating) 
\citep{sr98, wl03}, or momentum-based (i.e.\ pressure)
\citep{fce06,dqm10}.  \citet{co07} have included both mechanisms in
their ongoing work.  
Thus there are four conceptual components which feedback models may
take into account: mechanical-energy (a.k.a.\ bubbles),
mechanical-momentum (a.k.a.\ winds),
electromagnetic-energy (a.k.a.\ radiative heating), and
electromagnetic-momentum (a.k.a.\ radiation pressure).  The relative
importance of these components has been much debated among the authors just
mentioned.  A fifth component, the thin radio jets, emit comparable
amounts of energy and momentum with the other modes, but these intense
beams tend to drill through the galactic gas, depositing their energy
in the IGM. This makes jets less relevant as a feedback method, unless they
precess,  or there is a relative velocity between the AGN and
its environment, in which case significant heating of galactic gas
could occur \citep{ss09,sok09}.

There has naturally been much interest in reproducing the AGN feedback
process via numerical simulations. \citet{spring05} were among the
first to do so using a high-resolution 3D smoothed-particle hydrodynamics
(SPH) code.  They found that including an accreting BH particle which
returns energy to the surrounding gas reduces the fraction of baryons
which form stars, and at late times (for massive galaxies) expels most
of the remaining gas from the 
host, creating the classic ``red and dead'' elliptical.  Since then,
many more SPH simulations have been done, exploring various aspects of AGN
growth and feedback: \citet{pelu07}, \citet{khat08}, \citet{jbn09},
\citet{jnb09}, and 
\citet{mcc10}, to name a few.  

However, even though it has been shown that the emitted spectrum of
the average AGN power $EF_E$ has a strong peak in the X-rays
\citep[$\sim\!\!30$ keV:][]{sos04}, and moreover the ISM/IGM is known to be
optically thin to hard X-rays in most cases \citep{mm83}, all of the
simulations just 
mentioned include only the mechanical-energy component of feedback.  Some
recent studies have included other components: \citet{ost10} examine the
effect of mechanical-momentum feedback on AGN growth in 1 and 2D
simulations and find that including
this component reduces the final BH mass by a factor of 100;
\citet{dqm10} insert mechanical-momentum feedback into a 3D SPH code
(in fact the 
feedback was characterized as a radiation pressure, but only applied
to the central 0.2 kpc of the galaxies, making it functionally mechanical)  
to examine its effects on the major mergers of disk galaxies, and found
that with this component BHs self-regulate effectively during the
mergers, but without driving large quantities of gas out of the
galaxy altogether.  \citet{co07}, meanwhile, performed 1D simulations
with the full electromagnetic mode,
considering the heating and 
radiation pressure from the AGN radiation,  but without the mechanical
mode.  They studied the potential cooling flows of ``recycled'' gas
from dying massive stars, and found that the electromagnetic mode
alone was sufficient to drive out half of the incoming gas, with the
BH only accreting 1\% of the total (the remainder forming a
starburst).    However, 
until now no 3D simulation has been performed with the electromagnetic
mode included.

In \defcitealias{self09}{Paper~I}\citet{self09} (hereafter
\citetalias{self09}), we examined the 
effect of various ionizing radiation backgrounds on the properties of
massive elliptical galaxies.  In
\defcitealias{self10}{Paper~II}\citet[]{self10} (hereafter
\citetalias{self10}), we examined 
the same backgrounds with respect to the small satellite galaxies of
those massive ellipticals.  Now we turn to a different source of
ionizing radiation.  It is the
aim of this paper to present a first qualitative look at, and hint at
quantitative results from, an AGN feedback model incorporating an 
X-ray electromagnetic mode.

The paper is organized as follows.  In \S\ref{sect:bkg}, we describe
the numerical 
methods and parameters of our simulations, including the various AGN
feedback models.  In
\S\ref{sect:results} we 
describe the results obtained from those simulations, in particular
the effects of the different feedback models on the BHs themselves,
on the gas and stars in
the host galaxy and on the satellite galaxies.    Section~\ref{sect:disc}
is discussion and conclusion.

\section{Simulations and Parameters}\label{sect:bkg}

Our simulation code is GADGET-2, as in \citetalias{self10}, with the
only change being the addition of a black hole and associated
  feedback.  We use the UV background of \citet{fg09} (which was
  designated as ``FGUV'' in \citetalias{self09} and \citetalias{self10}). 
The simulation includes SN thermal feedback but not winds or any
stellar mass loss.  It
includes a simple prescription for 
metal-line cooling using cooling rates calculated by Cloudy \citep[v07.02,
last described in][]{cloudy}, which presumes photoionization and
collisional equilibrium for the gas and metal atoms, and assumes $0.1$
solar metallicity (note that our X-ray feedback code assumes solar
metallicity, as discussed in \S3.4).  A self-consistent treatment of
metallicity would increase the amount of cool gas available for
accretion at later times and
thus enhance the differences between our models. Our simulations  do not
include optical 
depth effects or radiative transfer, in particular the self-shielding
of dense star-forming 
regions from the ionizing backgrounds, although those regions would be
optically thin to X-rays regardless.  As in \citetalias{self10}, all
simulations 
were performed with initially $100^3$ each of SPH (i.e. baryon)
particles and DM 
particles, with a 
gravitational softening length of 
0.25 kpc for the gas and star particles and twice that for the dark
matter particles.  Gas and star particles have masses in the range
$4 - 7 \times 10^5 M_{\odot}$, depending on the size of the box; 
the assumed cosmology is
$(\Omega_M,\Omega_{\Lambda},\Omega_b/\Omega_M,\sigma_8,h)=(0.3,0.7,0.2,0.86,0.65)$ 
as in \citetalias{self10}.

The BH feedback works as follows.  At $z=9$ a single seed BH particle of mass
$1.5\times 10^6 M_{\odot}$ is created at the point of minimum
potential in the most massive progenitor halo. The seed mass is
chosen to roughly follow the Magorrian 
relation for the galaxies at this redshift; the AGN behavior is not
particularly sensitive to the seed mass \citep{hopkins06}.  The BH
grows at a modified Bondi-Hoyle-Littleton rate, 
\begin{equation}
\dot{M}_{\text{BH}}=\frac{4\pi\alpha_B (G M_{\text{BH}})^2 \rho}{(v_{\text{BH}}^2+c_s^2)^{3/2}},
\end{equation}
where $M_{\text{BH}}$ and $v_{\text{BH}}$ are, respectively, the mass
and speed (with respect to the surrounding gas) of 
the black hole,
and $\rho$ and $c_s$ are the density 
and sound speed of the gas in the SPH kernel centered on the BH
\citep{spring05,dsh05,hopkins06,jnb09}.  The 
free dimensionless parameter $\alpha_B$, which we choose to be $100$,
represents the fact that the gas density at the BH accretion radius is
likely to be much higher than the average density in the local SPH
kernel due to the limited resolution of the code
\citep{hopkins06,bs09}.  Future work could be done using the more
realistic density-dependent accretion efficiency of \citet{bs09},
which reduces to $\alpha=1$ for large accretion radii.  
The Bondi-Hoyle accretion is capped at the Eddington rate, 
$\dot{M}_{\text{Edd}}=4\pi G m_P M_{\text{BH}}/\epsilon_r c \sigma_T$.  When
the BH's notional mass as calculated from these accretion rates
exceeds its true dynamical mass, the BH particle swallows nearby gas
particles as needed to make up the difference.  We enable artificial
recentering of the BH particle on the potential minimum of the galaxy,
since with our modest mass resolution dynamical friction is
insufficient to keep the young BHs centered (though it is at late
times, when the recentering has no effect).                                 

We name our three feedback models ``BH'', ``BHX'' and ``BHXRP''.  The
``BH'' model consists only of thermal energy deposited at each
timestep in the SPH
kernel centered on the black hole,
in the amount of $\dot{E}=\epsilon_T \epsilon_r \dot{M}_{\text{BH}}
c^2$, where $\epsilon_r=0.1$ is the overall radiative efficiency of
the BH \citep{yt02}, and $\epsilon_T$ the fraction of the radiation
energy output which is assumed to be
absorbed thermally by the local gas.  Our choice of
$\epsilon_T=0.005$, together 
with our choice of $\alpha_B=100$, is designed to produce a final
($z=0$) BH mass roughly in line with the Magorrian relation for these
galaxies.  Note that our $\epsilon_T$ is 10 times smaller than than
the value used in other GADGET-2-based simulations
\citep[e.g.,][]{spring05, jnb09}, which is due to slightly different
star-formation criteria than the generic GADGET-2 (a lower density
threshold for star formation, a
star-formation timescale which is shorter at low densities but
constant at high densities and never shorter than the cooling time,
and requiring the gas to fulfill  a convergent flow criterion and the
Bonnor-Ebert criterion to form stars), the upshot of which is that our stars
form at somewhat lower 
gas densities, effectively lowering the BH accretion rate and thus its
mass.  Other simulations \citep[e.g.,][]{onb08, ost10} have used
similarly low efficiencies around $\epsilon_T = 0.01$---on the other
hand, the OverWhelmingly Large Simulations (OWLS) 
\citep{owls}, using a modified GADGET-3,  find 
$\epsilon_T=0.15$ to match the Magorrian relation, though also finding
that the value should be reduced 
for lower 
resolution \citep{bs09}. Furthermore, observational evidence suggests a value
 $\epsilon_T\approx 0.015$ \citep{mabk09}, although since we neglect
important processes like stellar mass loss and chemical evolution, our
value (and others') for $\epsilon_T$ is constrained far more by the
simulation than 
by physics.  
At any rate, we are primarily 
interested in the relative differences between the models with and
without radiative feedback, which
should not be significantly affected by the exact star-formation
prescription used.  We call the mechanical-energy component ``thermal''
feedback for short to 
distinguish it from the X-ray feedback below.  This energy component
 is of unspecified origin: if it is assumed to be the result of the broad-line 
wind, one should also include the momentum input \citep{ost10}, which
has typically been neglected and which will not be included here.  

The ``BHX'' model has the same feedback component just discussed, with
the same assumed efficiencies, but adds another: the
electromagnetic-energy component of X-ray
radiation from the AGN.  This radiation is emitted from the location
of the BH particle with a luminosity of $L_X=\epsilon_X \epsilon_r
\dot{M}_{\text{BH}} c^2$, with a bolometric-to-X-ray conversion term
$\epsilon_X$.
This luminosity is converted to a flux at each gas particle simply by
$F_X=L_X/4\pi r^2$, with $r$ the distance of the particle from the BH,
and the flux is converted to a heating rate for the gas using Eq. 36-43 in
\citet{co07} \citep[based on][]{sos05},  taking only terms which are
dependent on the ionization 
parameter $\xi\propto F_X$.  These equations are parametrized in terms
of bolometric flux, and implicitly assume $\epsilon_X\approx
0.04$ (as well as solar metallicity: see \S3.4).  We do not include radiative
transfer/optical depth effects for this radiation (again, the ISM/IGM 
generally has a low optical depth to X-rays). Nor do we include a
speed-of-light delay in propagation from the AGN across the box, but
since our box's high-resolution region is only 2 Mpc in radius and we
are most interested in the 
central 30 kpc, the effects of no delay should be small.  

The ``BHXRP'' model adds a third component: electromagnetic-momentum,
the radiation pressure from this 
X-ray flux, by applying to each gas particle a radial force away
from the BH particle equal to the X-ray heating rate at that timestep
divided by the speed of 
light $c$. We neglect
the effect of dust, which 
dominates the opacity by several orders of magnitude at temperatures
where it can exist ($T<10^{3.5}$~K) \citep{sem03}, making this component
perhaps an underestimate (although as discussed in \S 3.3 below, our
heating rate may be overestimated due to metallicity effects).  

As an additional point of comparison, we run a set of simulations
designated ``BH+X''.  These have the same feedback mechanism as the
regular BH runs, but also include an X-ray background field (which the
others do not).  This field is identical to the
``FGUV+X'' model in \citetalias{self10}; see that paper for
details. Briefly, the field peaks in strength around $z=2$ and falls
off sharply at earlier and later times, to roughly emulate the
observed quasar background (i.e., the X-ray heat due to galaxies other
than the one being simulated).    
This model allows us to differentiate the effects of X-rays originating in
local AGN feedback from generic (spatially uniform) background
radiation.  We also compare to a model with no AGN activity whatever,
``No BH''.  

We 
use three sets of initial conditions from \citetalias{self10} that
were designated 
galaxies/halos A, E, and M, and are so designated here as well.  These
are elliptical galaxies of virial mass $1-2\times10^{12}$ M$_\odot$.
 As will be discussed below, all the galaxies gave
the same qualitative results.  As in  
 \citet{naab07}, \citet{jno09} and \citetalias{self09}, we choose galaxy A as
 our fiducial 
 case, noting any differences from the others where relevant.  

Throughout the paper, all distances are physical except where noted,
with assumed $h=0.65$ as 
in \citetalias{self09} and \citetalias{self10}.

\section{Results}\label{sect:results}

\subsection{Evolution of the Black Hole Mass}
We first examine the effects of these various feedback mechanisms on
the black hole itself.  It is well-known that black holes are
characterized only by mass, electric charge, and angular momentum
\citep{bek98}, of which only the first is relevant here.  

\begin{figure}
\includegraphics[width=2.5in,angle=-90]{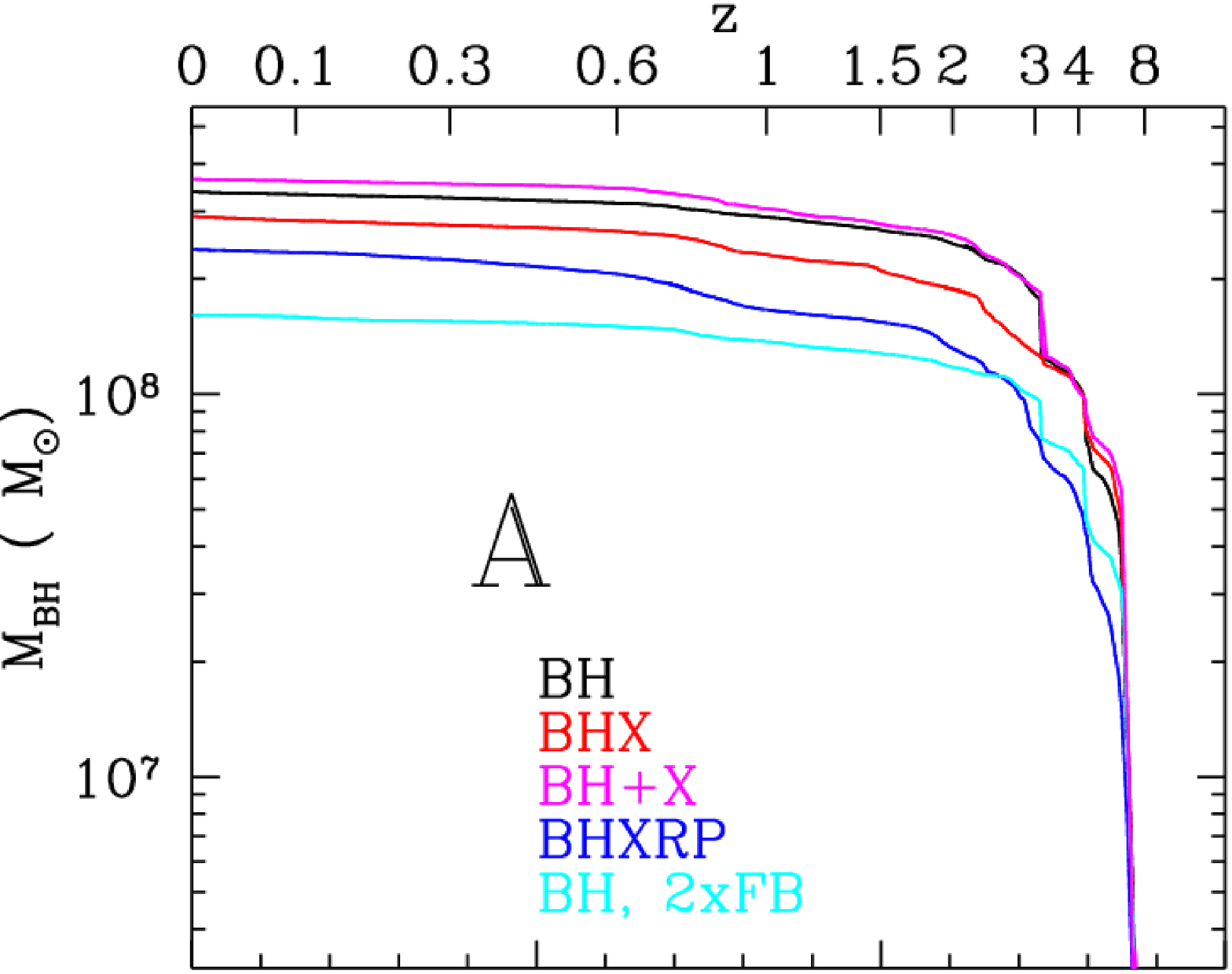}

\includegraphics[width=2.5in,angle=-90]{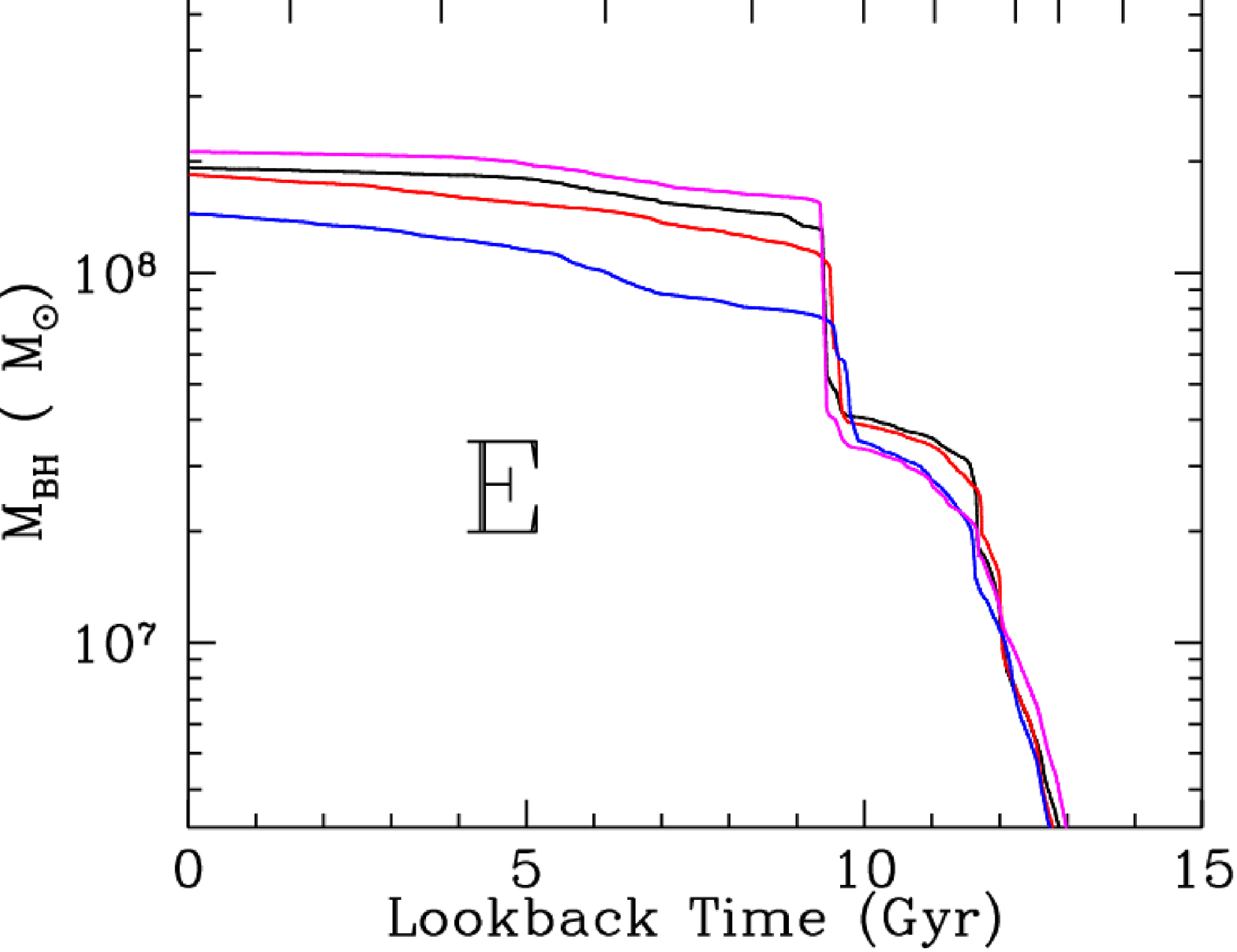}
\caption{The mass of the central black hole for our various feedback
  models, in galaxy A (top) and galaxy E (bottom).  Adding the
  X-ray feedback and radiation pressure decreases 
  the final BH mass by $30\%$, while increasing the thermal feedback
  by a factor of 2 reduces the final mass by a factor of 2.}
\label{fig:BHmass}
\end{figure}

The top panel of Figure~\ref{fig:BHmass} shows the mass of the black
hole for our four 
feedback models with galaxy A.  Also shown is a model where the
thermal feedback efficiency parameter $\epsilon_T$ has been increased
by a factor of 2 to $0.01$ (notated as ``BH, 2xFB'' in the figure).
We see that although the additional 
feedback present in the BHX and BHXRP models reduces the final black
hole mass by $15\%$ and $30\%$ respectively, compared to the BH
model, these changes are significantly less than the
factor of 2 reduction we see in the model with increased thermal
feedback---and another model with thermal feedback increased 10 times
has a lower BH mass by a factor of 25.  This is because the thermal
feedback directly affects only 
the immediately neighboring gas particles, which are the same
particles used to calculate the black hole's local density and thus
its accretion rate, whereas the X-ray feedback affects all gas
particles, thus making (as we will see) its effect on the host galaxy
relatively stronger than its self-regulation effect, compared to
thermal feedback. 

 On the other hand, the addition of an X-ray
background in the BH+X model \emph{increases} the final BH mass
slightly (9\%), due primarily to more gas being available in the
merger event at $z=0.7$.  
  All the
models except the enhanced thermal feedback model are consistent with
the Magorrian 
relation for this galaxy, which predicts a black hole mass of $\sim\!\!
2\times 10^8 M_{\odot}$, depending on model, using the fit of
\citet{hr04}---since the 
galaxies are ellipticals, we approximate ``bulge mass'' as the
stellar mass within three stellar half-mass radii ($\sim
18$~kpc) of the galaxy center.  We note 
here that if the electromagnetic feedback (heating and
radiation pressure) is used in the
absence of any mechanical feedback, the black hole becomes too large
by three orders of magnitude: it seems that for our simulations the
electromagnetic mode alone is insufficient for self-regulation.  This
differs from the result of \citet{co07}, who found effective
self-regulation using only electromagnetic feedback; we attribute the
difference to their inclusion of dust opacity, which is certainly very
important in AGN accretion regions \citep{sco03}, as well as their
neglect of infalling satellite galaxies and IGM.  

The lower panel of Figure~\ref{fig:BHmass} shows galaxy E, which is
much the same as galaxy A, except for the major (1:1 mass ratio)
merger event at  $z\approx 1.5$ (10 Gyr ago), which drives a huge
amount of gas to the BH and causes the mass to jump by a factor of
$\sim\!\! 5$.  Again in this case the BH+X model has the largest
final mass (by $\sim \!\!10\%$), thanks to the X-ray heating leaving more
gas available to 
be accreted in the merger event.  We also see that BHXRP accretes only
half the gas of the other models at the moment of merger, making up
the difference slowly over the next several Gyr; it seems that the
radiation pressure is effective in blowing out the merging gas when
X-ray heating and thermal feedback are not, consistent with \citet{dqm10}.
Otherwise, galaxies E and M are 
much the same as galaxy A, with BHX and BHXRP having lower black hole
masses by roughly $15\%$ and $30\%$.  

Another factor which we do not model here is the creation and mergers
of black holes in subhalos.  For simplicity, we have created only one
BH, in the largest halo at $z=9$ (which remains the largest halo until
the present).  While we do not expect the lack to change our results
qualitatively, including AGN in satellites would reduce gas mass and
star formation in those systems, leaving less gas and stellar mass to
be accreted by the central galaxy in minor and major mergers.  The
effect would be especially strong on systems like galaxies E,
with its
nearly 1:1 merger.  

\begin{figure}
\includegraphics[width=\linewidth]{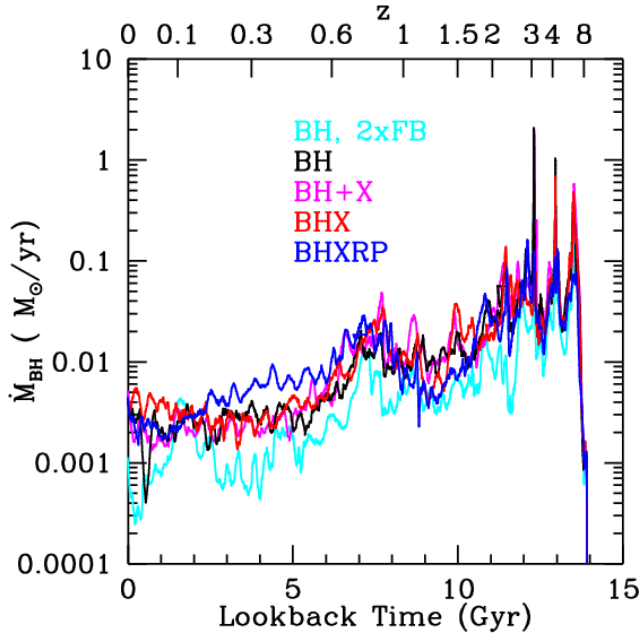}
\caption{The accretion rate for galaxy A in $M_{\odot}$/yr with the different
  feedback models.  Accretion peaks at $z=3$ and again with the major
  merger at $z=0.7$, after which BHXRP has a sustained period of
  higher accretion.   The X-ray luminosity is proportional to the
  accretion rate ($L_X=\epsilon_X \epsilon_r
\dot{M}_{\text{BH}} c^2$), 
  with $1 M_\odot/$yr corresponding to $2\times 10^{44}$ erg/s.
    The sharp spikes at $7>z>3$ correspond to bursts
  of Eddington-limited accretion.}
\label{fig:BHacc}
\end{figure}

Figure~\ref{fig:BHacc} shows absolute accretion rates for the same
 models with galaxy A.   The rates are smoothed over
160 Myr for clarity; the BH accretion shows high variability over
 timescales as short as a few timesteps ($\sim 100$ kyr). For the
 models with 
 lower thermal feedback, the 
rate peaks near $1 M_{\odot}$/yr around $4\gtrsim z \gtrsim 3$, and
slowly declines thereafter to a final value of $\sim\! 0.003 M_{\odot}$/yr,
except for the peak around $z\approx 0.7$, which corresponds to a
moderate ($6.5:1$) merger event for Galaxy A.  
 Of interest is
the suppression of the accretion spike at $z=3$ for the BHX model, and
all three accretion spikes (at $z=6,4,3$) by BHXRP: we see that these
feedback modes are effective at self-regulation for very high accretion
rates.  Also of note is that for
4~Gyr after the merger BHXRP has more accretion than
the other models by a factor of $2$ (which will be significant in
\S3.3).  

In the following subsections we disregard the model with enhanced thermal
feedback; in all cases its effects are the same or weaker than the
plain BH model.

\subsection{Impact on Gas}
\begin{figure}
\includegraphics[width=\linewidth]{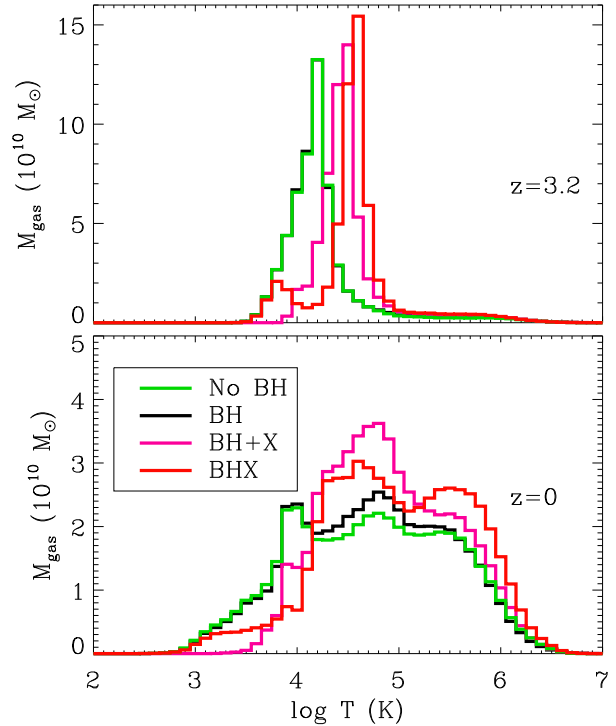}
\caption{Temperature distribution of all gas in the box for halo A at
  $z=3.2$ (top panel) and $z=0$ (bottom panel).  BHXRP, not shown, is
  nearly identical to BHX; No~BH and BH are nearly identical in the
  top panel.  The BHX and BHXRP
  models have a 
  10\% higher mean temperature than the BH+X model at $z=3.2$, and
  25\% higher at $z=0$.}
\label{fig:gasall}
\end{figure}

We now turn to the effects of the various BH models on the host
galaxy. We expect our X-ray feedback to be quite effective at heating
gas, and 
indeed it is.  Figure~\ref{fig:gasall} shows the temperature distribution
of all gas in the halo A box at $z=3.2$ (during the epoch of peak star
formation)  and the present.  At $z=3.2$, the No~BH model has the
coldest gas, with a mean temperature of $6.4\times10^4$~K. Adding BH
thermal feedback heats the gas $11\%$ to $7.1\times10^4$~K, and
adding an X-ray background heats it 24\% more to $8.8\times10^4$~K.
However the two BHX models are more effective by another 9\%, 
giving a mean gas temperature of roughly $9.6\times10^4$~K.  At
$z=0$, the effect of feedback X-rays is more pronounced compared to a
background: BHX and BHXRP, at a mean temperature of $3\times10^5$ K,
are 25\% hotter than BH+X, which in turn is only 14\% hotter than BH
and No~BH.  

\begin{figure}
\includegraphics[width=\linewidth]{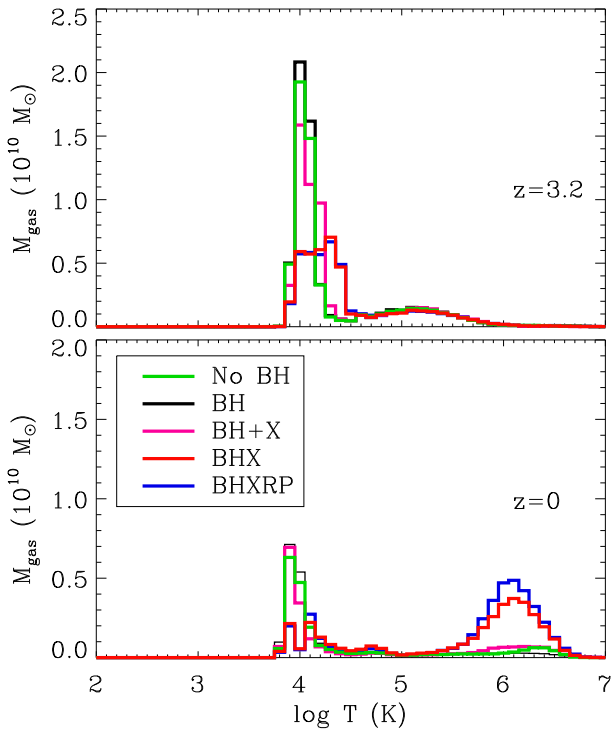}
\caption{Temperature distribution of virialized gas ($\rho>200 \bar{\rho}_b$) in
  the whole box for halo A at 
  $z=3.2$ (top panel) and $z=0$ (bottom panel).  At $z=3.2$, the X-ray feedback
  models have a  higher mean temperature than the other models by a
  factor of 1.5, and at $z=0$ they are hotter by a factor of 2.  }
\label{fig:gasvir}
\end{figure}

Figure~\ref{fig:gasvir} is the same as Figure~\ref{fig:gasall}, except
that it shows only virialized gas: gas with a density more than 200
times the mean baryon density.  Here the effect of the AGN feedback
X-rays is more pronounced, as we would expect, since most of the
virialized gas is near the central galaxy, where the AGN X-rays are
strongest.  At $z=3.2$, the three models without X-ray feedback are all
within $10\%$ of each other in mean temperature at roughly $9\times
10^4$~K, while BHX and BHXRP have mean temperatures of  $1.6\times
10^5$~K and  $1.3\times 10^5$~K respectively, 40-80\% higher than
the other models.  At $z=0$ the picture is even more extreme. BHX
and BHXRP have $2.4\times 10^{10} M_\odot$ and  $3.0\times 10^{10}
M_\odot$, respectively, of virial X-ray gas $> 10^{5.5}$~K, more than 4
times the amount that BH+X has, and 6 times the amount of the other
models.  Unlike the Warm-Hot Intergalactic Medium (WHIM), which is
usually defined as  $\rho < 100 
\bar{\rho}_b$ \citep{smith10}, this gas is dense enough to emit
significant soft X-rays.

We estimate the (Bremsstrahlung) X-ray luminosities of the
various models via
\begin{equation*}
L_X\propto \int \rho^2 \sqrt{T} dV
\end{equation*}
\citep{ev90} for gas above $2\times 10^6$ K over the virial volume (a 500 kpc
radius), and find that all models 
have $\log L_X 
\approx 38 - 39.5$, consistent with the gas luminosities found by
\citet{bkf11} for 30 early-type galaxies of similar size.  The
differences between BH models is modest: averaged over the
three ICs, the BH+X, BHX and BHXRP models have increased luminosity by
factors of 1.6, 2, and 3, respectively, compared with the models
without X-rays.  Adding  BH feedback also increases the X-ray
effective radius: only 33\% of the total X-ray luminosity for No~BH
comes from outside the central 10 kpc, while 63\% does for BH and
BH+X, 68\% for BHXRP and 73\% for BHX (again averaging the 3 ICs).  

\subsection{Impact on Stellar Mass}
\begin{figure}
\includegraphics[width=\linewidth]{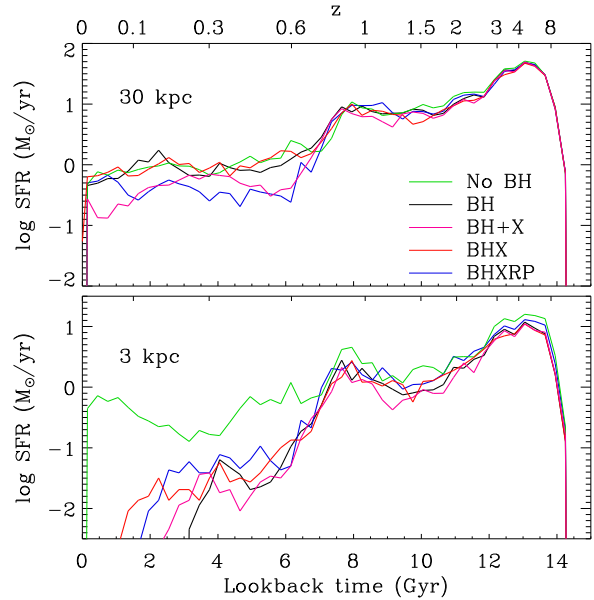}
\caption{Log of the star-formation rate over time for galaxy A with
  the various feedback 
  models, considering stars within 30~kpc (upper panel) and 3~kpc (lower
  panel) of the galaxy center at
  $z=0$.  Where the curves drop off the bottom of the plot they go all
  the way to zero.  BHXRP and BH+X are both effective at suppressing
  late star 
  formation for the whole galaxy, while in the center adding X-ray
  feedback actually keeps star formation going longer.  }
\label{fig:sfr-gal-A100}
\end{figure}

We naturally expect the hotter gas produced by the AGN X-ray feedback
to reduce the production of stars.  The upper panel of
Figure~\ref{fig:sfr-gal-A100} shows 
the star-formation rate (SFR) over time for galaxy A, out to a radius of
30 kpc.  We see only modest differences in the initial star-formation peak
(see below),
but BH+X and BHXRP are both effective at suppressing late star
formation: BHXRP has a lower SFR than the other models by 0.5~dex for
the the last 6 Gyr ($z<0.6$), while BH+X suppresses the SFR by 0.8 dex
for the last 1.5 Gyr ($z<0.1$).  BHXRP forms $3.4\times10^9$ M$_{\odot}$
of stars in the host galaxy
after $z=0.5$, and BH+X $3.6\times10^9$ M$_{\odot}$, which is less than
half of the roughly 
$7.2\times10^9$ M$_{\odot}$ formed by the other models.  The BH+X result
is interesting in the 
context of 
\citetalias{self09}, where the X-ray background produced a strong
burst of late star-formation, as gas that had been kept hot through
the X-ray background peak at $z\approx 2$ finally cools and flows to
the center of the galaxy to form stars (the so-called ``cooling
flow'').  Here, a cooling flow seems to 
be forming around redshift of 0.3 (see the lower panel of
Figure~\ref{fig:sfr-gal-A100}), but the AGN effectively shuts it off.  The 
effect with BHXRP, meanwhile, is clearly related to its enhanced
accretion rate after the merger event: we see from
Figure~\ref{fig:BHmass} that BHXRP accretes less gas during a major
merger, leaving a residual which forms a few central stars (see
below), but more 
importantly powers extra feedback for the next several Gyr,
suppressing star formation at larger radii.    

The lower panel of
Figure~\ref{fig:sfr-gal-A100} shows 
the SFR  for galaxy A out to a radius of only 3 kpc (the 3D
stellar half-mass radius for the galaxy is the range $5-8$~kpc). The
results here 
are somewhat different: while all the models with BHs have no star
formation at the present, BHX and BHXRP have nonzero star formation
until roughly 1.5 Gyr ago, compared to 3 Gyr for the plain BH model
(and 2.5 Gyr for BH+X).  In terms of mass, BHX forms $1.7\times10^8$
M$_{\odot}$ in the central 3 kpc
after $z=0.5$, BHXRP $2.8\times10^8$ M$_{\odot}$ (the residual
from the merger just discussed), which is $2-3$
times more than the $1\times10^8$ M$_{\odot}$ formed by BH and BH+X, but
still less than a tenth of the No BH model ($3.1\times10^9$ M$_{\odot}$).
 This is an expected result: as in \citetalias{self09}, when  X-ray
 heating is present, more gas is available at moderate and low
 redshift ($z<2$) to be drawn to the
 galaxy's center in major mergers and in cooling flows, and the AGN
 feedback, while eventually preventing most of this gas from forming stars,
 cannot do so immediately.  

\begin{deluxetable}{llcccc}
\tablecolumns{6}
\tablewidth{0pt}
\tablecaption{Stellar mass results
\label{tab:stars}}
\tablehead{
\colhead{IC}&\colhead{Feedback} & \colhead{$M_R(5\text{kpc})$}&
\colhead{$M_R(30\text{kpc})$}& \colhead{$M_R(2\text{Mpc})$} & \colhead{$\epsilon_\star$}}
\startdata
A & No BH & 9.58 & 19.43 & 50.35 & 0.543\\
A & BH & 6.27 & 18.12 & 48.09 & 0.525\\
A & BH+X & 5.85 & 16.62 & 42.70 & 0.476\\ 
A & BHX & 6.46  & 17.05 & 42.99 & 0.472\\
A & BHXRP & 7.70 & 17.76 & 43.90 & 0.481\\
E & No BH & 8.59 &15.53  &38.61 & 0.580 \\
E & BH &7.45  &14.70  &37.98 & 0.534\\
E & BH+X &7.41  &14.50 &34.39 &0.536\\ 
E & BHX & 8.16  &14.48 &34.00 &0.521\\
E & BHXRP &7.48  &14.64 &33.29&0.514\\ 
M & No BH &6.64  &11.91  &25.72&0.631 \\
M & BH &5.42  & 11.91 &23.64&0.594 \\
M & BH+X & 4.62 &10.02 & 20.64&0.561\\ 
M & BHX & 5.34  &10.77 & 20.06&0.596\\
M & BHXRP & 5.51 &10.22 & 20.21&0.553\\ \enddata
\tablecomments{Results are for stars within the specified radius of
  the principal halo at $z=0$.  Masses are $10^{10}M_\odot$.  The last
column is the ratio of stellar to total (implied) baryonic mass for
the central galaxy. Adding X-rays significantly decreases the total
stellar mass and galactic stellar mass, while the BHXRP model actually
increases the mass in the 
center of the host galaxy, 
and $\epsilon_\star$,
in 2 of 3 cases. }
\end{deluxetable}

\begin{figure}
\includegraphics[width=\linewidth]{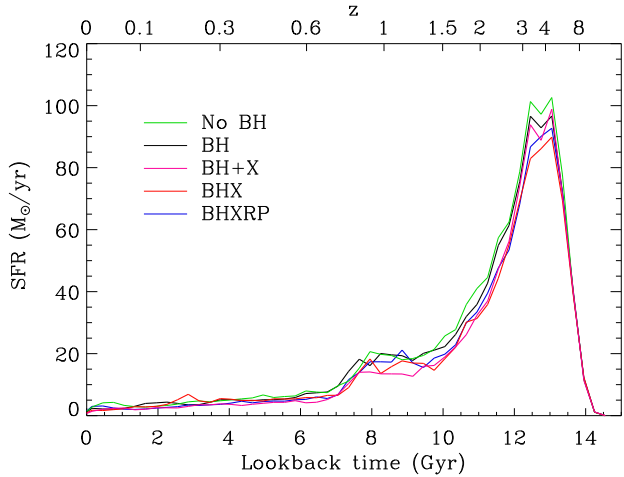}
\caption{Star-formation rate over time for the various feedback
  models, considering stars within the whole box (2~Mpc radius) of
  galaxy A (galaxies 
  E and M are similar).    BHX and BHXRP
  effectively suppress the SFR peak at $4>z>3$, while the X-ray
  background reduces star-formation at $1.5>z>1$. }
\label{fig:sfr-big}
\end{figure}

Table~\ref{tab:stars} shows the stellar mass of the host galaxy for
the various models and ICs at 
three radii: 5 kpc, 30 kpc and 2 Mpc (which is essentially the whole
box).  The total star formation in the box (column 5) is suppressed
slightly (5\%) 
by the presence of the AGN, but the X-rays, whether  from the AGN or
the background, reduce the total stellar mass by 15\%,
three times as much. However, the background X-rays and the
feedback X-rays cause their suppression differently: as we see
 in  Figure~\ref{fig:sfr-big}, which shows the star-formation history
 of the entire box for the halo A runs, the two models with
 X-ray feedback suppress the star-formation peak at $4>z>3$, while
 BH+X is most effective at a somewhat lower redshift, $1.5>z>1$.  
 This clearly reflects the relative timing of the X-ray flux
 between the feedback and background
sources, since the X-ray background peaks in intensity at $z\approx2$,
while the X-rays from feedback peak when the BH accretion rate does,
at $4>z>3$. Thus BHX can have a strong effect on the SFR peak---recall
that in Figures~\ref{fig:gasall}~\&~\ref{fig:gasvir} we saw that BHX
has hotter gas than BH+X at $z=3.2$---while later  when the local AGN
is less active, the X-ray background is 
relatively more effective.   
This difference in timing also produces a substantial effect on how
the stellar mass is distributed, as we will 
 see in the next
 subsection.

The stellar mass in the central galaxy (column~4 of
Table~\ref{tab:stars}) is more or less as one would expect: thermal
feedback and X-rays each reduce the stellar mass by a modest amount
($\sim \!\!10\%$). 
To present the same results in a different way, the last column of
Table~\ref{tab:stars} shows the baryon-conversion efficiency for the host
galaxy, in this case identified using the Amiga Halo Finder
\citep[AHF:][]{kk09}.  The value $\epsilon_\star$ is defined as it was
in \S4.2 of 
\citetalias{self10}, $\epsilon_\star \equiv
M_\star/M_\text{Halo}(\Omega_M/\Omega_b)= 5 M_\star/ M_\text{Halo}$ 
for our cosmology, where $M_\star$ is the halo's mass in star
particles within $0.1$ virial radii of the center and $M_\text{Halo}$
its total virial mass.  As discussed 
in \citetalias{self10}, the absolute 
efficiencies for our models are all significantly higher than the value
derived from SDSS by \citet{guo10}, $\sim \!0.2$, but we are interested
in the differences between the models.  The two models with only
thermal feedback reduce the efficiency by $0.035$ from the No BH case, while
adding the X-ray feedback (or an X-ray background in 2 of 3 cases)
reduces it by an 
additional $0.035$.  So our 
most effective model, BHXRP, reduces the baryon-conversion efficiency by
7\% from No BH.  This is only one fifth of the reduction which would be
necessary to bring our simulations in line with \citet{guo10}, but
still a significant difference.  

\begin{figure}
\includegraphics[width=\linewidth]{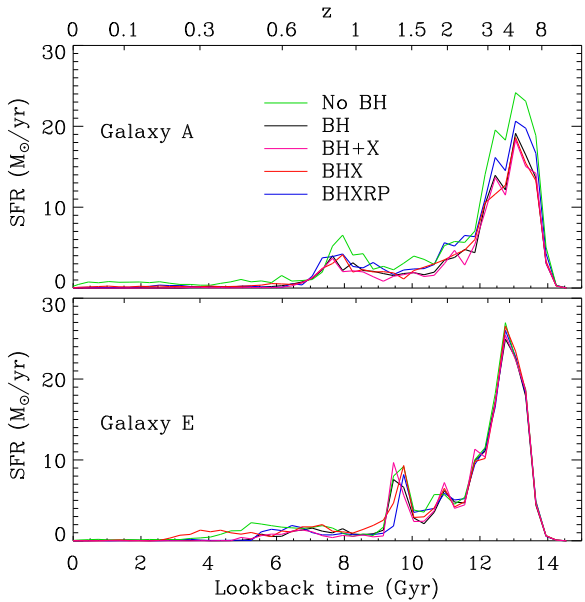}
\caption{Star-formation rate over time for the various feedback
  models, considering stars within 5 kpc
  of the galaxy center for galaxy A (top panel) and galaxy E (bottom panel) at 
  $z=0$. In galaxy A,  BHXRP allows significantly more central star formation
  than the other BH models, while the major merger in galaxy E causes
  more star formation for BHX. }
\label{fig:sfr-small}
\end{figure}

We find more unexpected effects in column~3 of
Table~\ref{tab:stars}, which gives the stellar masses for the central
5 kpc of
the galaxy. For galaxy A, the BHXRP model has almost 20\% more mass than the
other AGN models (though still 20\% less than the No BH case), and 4\%
more than BHX at the 5~kpc radius.  Once
again this turns out to be associated with the SFR peak (though a
small part comes from the extra late star formation discussed above):
we see in the 
upper panel of  Figure~\ref{fig:sfr-small} that BHXRP has significantly
more star formation in the peak at $z\approx 4$.  This is clearly
related to what we saw in the
previous subsection: since BHXRP's feedback is more effective at self
regulation, it accretes less strongly at $z\approx 4$ and thus
provides less heating to the surrounding gas, which can thus form more
stars.  I.e., since the gas is pushed out by the feedback instead of
merely being heated, the ability of the gas to form stars is less
impaired even as the BH reduces its own growth rate.  This effect
is not robust, however: in the smaller galaxy M BHXRP has only barely
more central mass than the other BH models, and in galaxy E it is BHX
which has the most.  But in this case BHX has slightly less stellar
mass within 30 kpc than BHXRP, so it is only the central concentration
which has been enhanced.  This is related to the major merger event
which galaxy E experiences: the extra residual gas from the
early X-ray heating compared the no-BHX models means more gas is driven
to the center to form stars during and after the merger, which we can
seen in the lower panel of Figure~\ref{fig:sfr-small}.   Meanwhile the
radiation pressure in BHXRP forces this gas back out, as suggested by
Figure~\ref{fig:BHmass}, so it forms 
stars at larger radii, in agreement with \citet{dqm10}. 

\begin{figure}
\includegraphics[width=\linewidth]{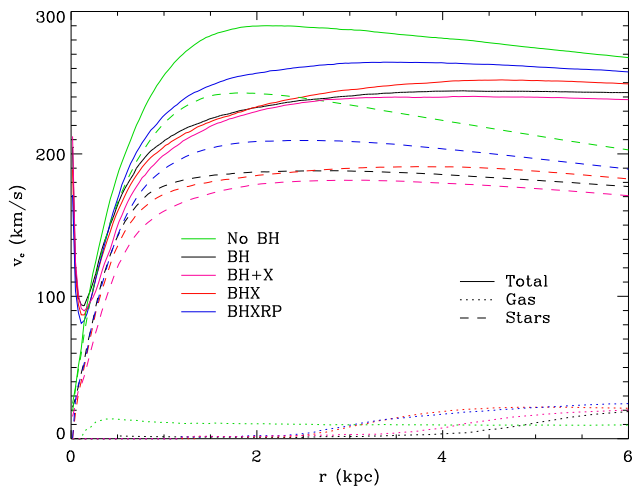}
\caption{Circular speed for stars (dashed line), gas (dotted line), and total
  (solid line; also includes dark matter and the BH particle), for
 galaxy A with the various feedback models, 
    $z=0$.
   The upturn for the total circular speed curves in the
  innermost 200~pc is due to the mass of the BH particle itself
  (hence it is not present in the No~BH model).  Adding BH feedback
  lowers the peak value and flattens 
  the inner slope, although adding radiation pressure reverses those
  effects somewhat for this galaxy.  All BH models remove residual gas
  from the inner 2 kpc.} 
\label{fig:vcirc-A100-z0}
\end{figure}

Figure~\ref{fig:vcirc-A100-z0} gives the radial mass distributions for
galaxy A in the form of circular speeds.  As we would expect from  column~3 of
Table~\ref{tab:stars}, No BH has the highest peak speed and the
steepest inner slope, with BH, BH+X and BHX lower by about 25\% in
peak stellar circular speed and BHXRP in between.  We can also
parametrize the mass distribution with a ``Faber-Jackson'' statistic,
$\sigma^4/M_\star$, where $\sigma$ is the (3D) stellar velocity dispersion
and $M_\star$ the stellar mass within some radius of the galaxy center (we
choose 30 kpc).  This statistic should be roughly constant for various
elliptical
galaxies \citep{fj76}, but we find a variation of 40\% between models,
with the No BH case being the highest at $0.065$
(km/s)$^4$/M$_{\odot}$ (corresponding to the most central
concentration),  BH+X and BHX the lowest at $0.047$ 
(km/s)$^4$/M$_{\odot}$, and the other models intermediate.  Similar
results are obtained with the other galaxies.   All these
values are lower by a factor of 
$\sim\!3$ than the
observational value of $0.22$ (km/s)$^4$/M$_{\odot}$ derived from the
$M_\star - \sigma$ relation compiled by 
\citet{rob06}, but within the
intrinsic scatter (and quite similar to those authors' simulation
results for $z=0$).  

\subsection{Effects beyond the host}
Inspired by the differences between BH+X and BHX in their overall
star-formation histories 
(Figure~\ref{fig:sfr-big}), we revisit the topic of
\citetalias{self10} and examine the low-mass slope of the galaxy mass
spectrum, as parametrized by the $\alpha$ of \citet{s76}, which we will call
$\alpha_S$ to distinguish it from the Bondi accretion parameter
$\alpha_B$.  As in \citetalias{self10} $\alpha_S$ 
is defined for our purposes by $n(M)dM\propto (M/M_*)^{\alpha_S}
e^{-M/M_*}dM$.  The results are given in Table~\ref{tab3:sch}: while
the No~BH and BH models agree within error 
with the values
corresponding to their background from
\citetalias{self10} (and BH+X has a modestly steeper slope likely
related to the suppression of cooling flows), the BHX
and BHXRP models indeed have an extreme effect on $\alpha_S$, flattening
it well beyond the observed value of $\sim\! -1$ to $-0.44$ and $-0.55$,
respectively. In \citetalias{self10} we found that $\alpha$ increases
as any heating is added: from $-1.6$ with no heating, to $-1.3$ with a
UV background, to $-1.0$ with UV and SN feedback, to $-0.75$ with UV, feedback
and an X-ray background.  As mentioned above, the timing of the
X-rays seems to explain the difference between BH+X and BHX, since the BHX
X-rays peak just before the epoch of primary  star formation in
low-mass galaxies 
(found in \citetalias{self10} to be $3\gtrsim z \gtrsim 2$), while the
X-ray background peaks toward the end of it, and doesn't significantly
suppress star formation until $z\approx 1.5$.  

This result suggests that our X-ray feedback might be too strong:
in fact, our formula converting X-ray flux to heating rate
has a term (representing the photoionization heating and line and
recombination cooling, Eq.~A35 in \citealt{sos05}) which is linear in
$Z/Z_{\odot}$---i.e. the 
metallicity as a fraction of solar---where solar metallicity is
assumed (Sergey Sazonov, private communication).  Thus, since the
metallicity of the ISM could be $0.1 Z_{\odot}$ at early times, our
heating rate could be too high by that factor.  
Moreover, the work of \citet{hh03} suggests that after reionization
the equilibrium temperature of the IGM is roughly independent of the
intensity of the radiation field, depending only on its spectrum, so
our $r^{-2}$ attenuation factor may be less significant than we
would na\"\i vely think.  

The results for $f M_{\text{crit}}$ are more modest.  The quantity $f
M_{\text{crit}}$ is defined in \citetalias{self10}: in short, it
represents the largest halo mass at which halos have an average
star:DM mass ratio of less than half the global value ($0.08$).  ($
M_{\text{crit}}$  is the theoretically-calculated virial mass whose escape
velocity is 
equal to the sound speed of its gas at the epoch when it should be
forming stars; the effective correction factor $f\approx 0.75$.) Here
again, No BH, BH, and 
BH+X agree well with the values that \citetalias{self10} gives from
their ionizing background models. The  X-ray background models have
somewhat higher values, as we would expect from their flatter low-mass
slopes.

\begin{deluxetable}{lcc}
\tablecolumns{3}
\tablewidth{0pt}
\tablecaption{Schechter $\alpha$ values and maximum ``barren'' halo
  masses with various backgrounds
\label{tab3:sch}}
\tablehead{
\colhead{Name} & \colhead{$\alpha_{\star}$} 
& \colhead{$\log f M_{\text{crit}}$}  }
\startdata
No BH & $-1.09 \pm 0.04$ & $8.67\pm 0.07$  \\
BH & $-0.96 \pm 0.14$ & $8.65\pm 0.01$  \\
BH+X & $-0.84 \pm 0.12$ & $8.76\pm 0.09$ \\
BHX & $-0.44\pm0.13$ & $8.95\pm 0.08$\\
BHXRP & $-0.55 \pm 0.08$ & $8.88\pm 0.03$\\ \enddata
\tablecomments{The Schechter-$\alpha$ values for the star
  particles, maximum ``barren''
  halo masses, and overall baryon-conversion efficiency with various
  ionizing background and AGN feedback models.  X-ray feedback from
  the AGN greatly suppresses small stellar systems.  }
\end{deluxetable}

\section{Discussion \& Conclusions}\label{sect:disc}
The effects of X-ray feedback from AGN are manifold.  We find that
 X-ray heating and radiation pressure are only moderately effective at
 self-regulation: they reduce the black hole's
mass  far less than  increasing the thermal feedback
 efficiency does, primarily by suppressing bursts of Eddington-limited
 accretion at early times.  The model with radiation pressure also
 accretes significantly less gas at the time of a major merger,
 instead accreting it more smoothly over the following several Gyr.
 The X-ray feedback produces 
 a significant reduction in the host galaxy's baryon-conversion
 efficiency compared to a traditional feedback model, but only
 slightly more than a model with traditional feedback and an X-ray
 background.  We note however that our baryon-conversion efficiency
 remains well above the observationally-derived value; though this
 problem is hardly unique to the present work it remains troublesome.
 On the other hand, less star formation would leave more gas available
 for AGN accretion, which would likely enhance the relative effect of
 electromagnetic feedback.

 The enhanced accretion and associated feedback in the
 radiation-pressure model can also 
 sustain a half-decade reduction  in the star-formation rate of the
 host galaxy for
 several Gyr after a major merger event, although the gas required for
 this extra feedback leads to more star formation in the central
 regions, which in turn can lead to enhanced  central concentration. 
 The AGN X-ray feedback also
 produces a significant mass of virialized, soft-X-ray-emitting gas at the
 present, which the X-ray background does not have (when ordinary AGN
 thermal feedback is present; in \citetalias{self09} we found a
 significant mass in hot, dense gas for a model with X-ray background
 but no AGN feedback), which increases both the X-ray luminosity and
 the X-ray half-light radius.   In a
serendipitous final result, we 
find that this X-ray feedback is also much more effective than an
X-ray background in suppressing
small galaxies and thus flattening the  low-mass slope of the galaxy
mass spectrum, due to its feedback's local origin making it effective
 at heating 
 gas some $2-3$~Gyr earlier than the background.  

Since AGN are known to emit X-rays through their host galaxies, we
view this study as a vital first step toward a more complete model of
AGN feedback. Moreover, since the X-ray luminosity of AGN 
is relatively well constrained
by observations (though not without intrinsic scatter), there is little
need for a new
free parameter to join the current 
$\alpha_B$ and $\epsilon_T$ (modulo the effects of metallicity and
dust).  Thus we hope that this ``new'' feedback mode will be employed in
future SPH simulations of AGN, since it is both undeniably present
and, as we have shown, substantial in effect.


JPO was supported by NSF grant AST 07-07505 and
NASA grant NNX08AH31G.  TN and PHJ acknowledge support by the 
DFG cluster of excellence `Origin and Structure of the Universe'.

\end{document}